# Bloch phonon-polaritons with anomalous dispersion in polaritonic Fourier crystals

Sergey G. Menabde,[1] Yongjun Lim,[2] Alexey Y. Nikitin,[3] Pablo Alonso-González,[4] Jacob T. Heiden,[1] Heerin Noh,[5] Seungwoo Lee,[2,5,6] and Min Seok Jang[1,*]

[1] School of Electrical Engineering, Korea Advanced Institute of Science and Technology, Daejeon 34141, Korea
[2] Department of Biomicrosystem Technology, Korea University, Seoul 02841, Korea
[3] Donostia International Physics Center (DIPC), Donostia-San Sebastián 20018, Spain; IKERBASQUE, Basque Foundation for Science, Bilbao 48013, Spain
[4] Department of Physics, University of Oviedo, Oviedo 33003, Spain; Center of Research on Nanomaterials and Nanotechnology, CINN (CSIC-Universidad de Oviedo), El Entrego, Spain
[5] KU-KIST Graduate School of Converging Science and Technology, Korea University, Seoul 02841, Korea
[6] Department of Integrative Energy Engineering, Korea University, Seoul 02841, Korea
*jang.minseok@kaist.ac.kr

**ABSTRACT:** The recently suggested concept of a polaritonic Fourier crystal (PFC) is based on a harmonically-corrugated mirror substrate for a thin pristine polaritonic crystal layer. The propagating polaritons in PFC experience a harmonic and mode-selective momentum modulation leading to a manifestation of Bloch modes with practically zero inter-mode scattering. PFC was first demonstrated for the hyperbolic phonon-polaritons in hexagonal boron nitride (hBN) within its Type II Reststrahlen band (RB-II) where the in-plane components of the dielectric permittivity tensor are isotropic and negative, while the out-of-plane component is positive. By contrast, a Type I Reststrahlen band (RB-I) is characterized by negative out-of-plane and positive in-plane permittivity components, and consequently, the inversion of field symmetry of phonon-polaritons compared to RB-II. Behavior of such RB-I modes in a polaritonic crystal is yet to be explored. Here, we employ a biaxial crystal alpha-phase molybdenum trioxide (α-MoO$_3$) and near-field imaging to study polaritonic Bloch modes in a one-dimensional PFC within the RB-I where the mid-infrared phonon-polaritons in α-MoO$_3$ have anomalous dispersion and negative phase velocity. Surprisingly, we observe a manifestation of Bloch waves as a dispersionless near-field pattern across the first Brillouin zone, in contrast to RB-II case demonstrated with in-plane isotropic hBN. We attribute this difference to the opposite field symmetry of the lowest-order phonon-polariton mode in the two RBs, leading to a different momentum modulation regime in the polaritonic Fourier crystal. Our results reveal the importance of mode symmetry for polaritonic crystals in general and for the emerging field of Fourier crystals in particular, which promise new ways to manipulate the nanolight.

**MAIN TEXT:** When polaritons propagate in a periodic environment with the period comparable to their wavelength, the dispersion of these quasiparticles may experience bending, potentially forming flat bands and opening a polaritonic bandgap[1]. Recently, the booming field of van der Waals (vdW) crystals has provided a wide range of low-dimensional polaritonic materials that can be exfoliated into thin flakes, sometimes as thin as a single atom[2-4]. Polaritons in such thin waveguides are able to confine light within nanometer-thick layers – orders of magnitude smaller than the free-space wavelength[5,6]. Naturally, creating a polaritonic crystal for modes in vdW materials is of great importance for a wide range of future applications that include ultra-compact modulators, filters, or cavities for enhancing Purcell effects.

Several schemes have been proposed to realize a polaritonic crystal with vdW materials, all of which require nano-patterning of either the substrate[7-10] or the polaritonic waveguide[11-15]. However, such binary structures inevitably contain sharp material edges that scatter the nanoscale-confined electromagnetic fields. For example, the edge-mediated intermode scattering of phonon-polaritons (PhPs) is leveraged in so-called hypercrystals[10,15], where the band diagram is filled with multiple high-order modes supported by a polar dielectric slab within its Reststrahlen bands (RBs).

In contrast, the recently proposed concept of a polaritonic Fourier crystal (PFC) is based on a harmonically-corrugated metallic mirror substrate and a pristine polaritonic material[16], and thus, advantageously, is free from sharp material edges. Furthermore, the smoothly varying geometry of the mirror provides a harmonic and mode-selective modulation of the polariton momentum due to its strong dependence on the distance between the waveguide and the mirror.

At the same time, the two most studied polaritonic vdW materials – hexagonal boron nitride (hBN)[17,18] and alpha-phase molybdenum trioxide ($\alpha$-$MoO_3$)[19,20] – have been employed to create prototypes of polaritonic crystals, as they support relatively low-loss hyperbolic PhPs at mid-infrared frequencies. While hBN is a uniaxial crystal with isotropic in-plane permittivity, $\alpha$-$MoO_3$ is a biaxial crystal. Up to now, polaritonic crystals based on hBN and $\alpha$-$MoO_3$ have been studied only within the Type II RB (RB-II) where at least one of the in-plane permittivity components is negative and the out-of-plane component is positive[10-14].

Here, we investigate the PFC based on $\alpha$-$MoO_3$ within the frequencies of Type I RB (RB-I) where the out-of-plane permittivity component is negative and both in-plane components are positive[19-23]. In this frequency band, PhPs exhibit negative dispersion (and as a result, negative phase velocity). Our near-field experiments reveal a different behavior of PhPs in the PFC operating in RB-I compared to our initial study conducted in RB-II using hBN[16]. We observe a manifestation of the Bloch mode across the whole first Brillouin zone, in the regime when the PhP wavelength, $\lambda_P$, is between $P$ and $2P$, where $P$ is the period of the PFC. Our results demonstrate how the field symmetry of the propagating PhPs in different RBs could affect their behavior in polaritonic crystals in general, and its particular importance for the emerging field of the Fourier crystals.

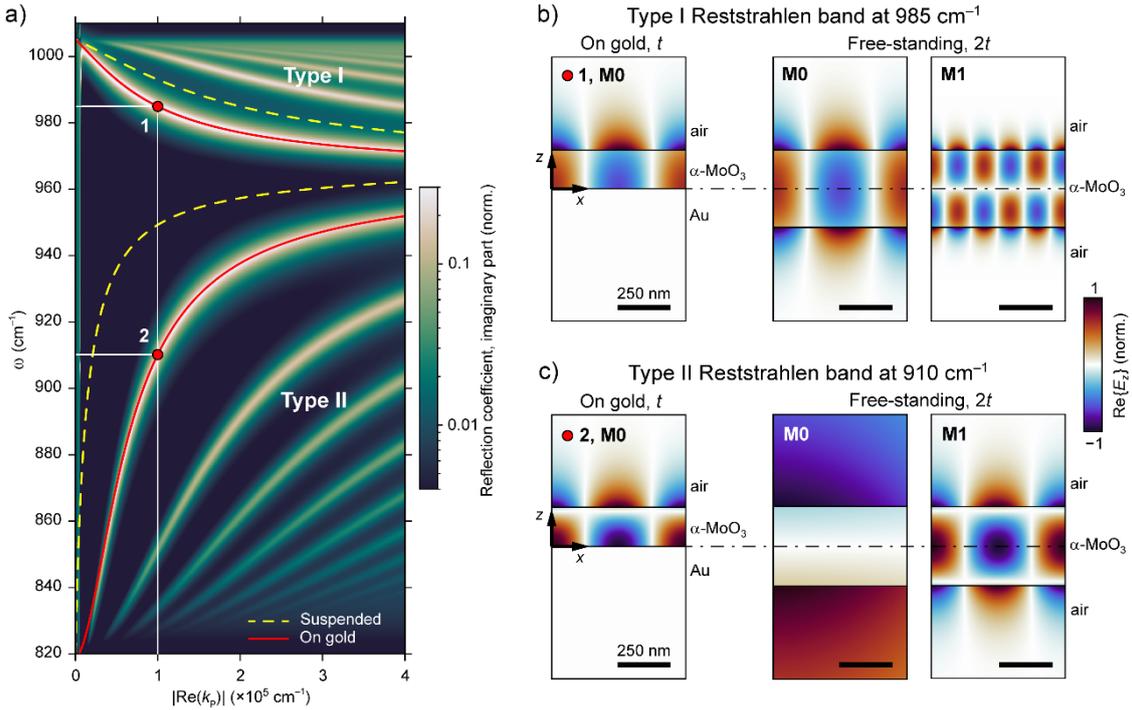

**Figure 1.** (a) Calculated dispersion of hyperbolic phonon-polariton modes supported by a 185 nm-thick α-MoO$_3$ on gold substrate (color map) within the RB-I and upper RB-II, both along the {100} crystalline direction. Also shown is the analytically calculated dispersion of the M0 PhP mode (red solid) and the dispersion of the M0 mode in free-standing α-MoO$_3$ of the same thickness (yellow dashed). (b) Calculated spatial distribution of the out-of-plane electric field of the first-order mode in 185 nm-thick α-MoO$_3$ on gold (left), and that of the M0 and M1 modes in the free-standing α-MoO$_3$ of double thickness (right); all at $\omega$ = 985 cm$^{-1}$ within the RB-I. Note that first-order mode is identical in both cases. (c) Same as in (b), calculated at $\omega$ = 910 cm$^{-1}$ within the RB-II. Here, the first-order mode on gold (left) is identical to the second-order mode in the free-standing and twice thicker α-MoO$_3$ (right). Calculation frequencies in (b) and (c) are such that the first-order mode on gold in both RBs have the same momentum, as indicated by the white guidelines in (a). All field plots are in the same scale.

In general, thin slabs of polar vdW dielectrics support a very large number of propagating PhP modes (equal to the number of monolayers in the slab). Their dispersion can be numerically obtained by mapping the imaginary part of the Fresnel reflection coefficient in momentum space using the transfer matrix method[24], or analytically calculated under the high-momentum approximation[25] $k_P \gg k_0$, where $k_P = 2\pi/\lambda_P$ is PhP momentum and $k_0$ is that of free-space light.

The multitude of PhP modes supported by a 185 nm-thick α-MoO$_3$ flake on a gold substrate is exemplified in Fig. 1(a) (color map), calculated for the frequencies of the RB-I ($\omega$ = 964–1005 cm$^{-1}$) and the upper RB-II ($\omega$ = 820–965 cm$^{-1}$) along the {100} crystalline axis[19-23]. The dielectric function of α-MoO$_3$ used for calculations is provided in the Supporting Information. Figure 1(a) also shows the analytically calculated dispersion of the fundamental (lowest-order) PhP mode (M0) in α-MoO$_3$ on gold (solid red) and when it is free-standing (yellow dashed). Note the opposite dispersion trend in RB-I and RB-II when α-MoO$_3$ is lifted from the gold substrate (free-standing α-MoO$_3$): M0 momentum becomes much smaller in RB-II, while it increases in RB-I. This important observation can be understood by considering the field profile of M0 PhP mode in both RBs.

First, we plot the calculated out-of-plane component of the electric field, $E_z$, of the M0 PhP mode in α-MoO$_3$ on gold at two different frequencies within RB-I and RB-II. For clarity, we consider the frequencies where M0 has the same momentum $k_{P,M0} = 1\times10^5$ cm$^{-1}$ in both RBs: 910 cm$^{-1}$ and 985 cm$^{-1}$, indicated by the white guidelines in Fig. 1(a). The $E_z$ distributions along the single oscillation are shown in the left panels of Figs. 1(b,c). In both cases, $E_z$ is screened by the image charges in the metal, forming an image phonon-polariton[21,26,27]. Due to the gold surface acting as a perfect electric conductor-like mirror, the screened M0 PhP mode is identical to a certain PhP mode in a twice thicker free-standing waveguide which has $E_z$ distribution with mirror symmetry[27]. We illustrate this by plotting $E_z$ for the M0 and the second-order (M1) PhP modes in the twice-thicker free-standing α-MoO$_3$ in both RBs (right panels in Figs. 1(b,c)). All panels in Figs. 1(b,c) are plotted in the same scale, and the gold surface is aligned with the symmetry plane of the twice-thicker slab ($z$ = 0; shown by the dash-dotted line). Notably, M0 and M1 PhP modes in the free-standing waveguide have the opposite field symmetry with respect to the plane $z$ = 0 in different RBs. Consequently, when mirrored by gold, only the mode with symmetric $E_z$ is allowed as it has $E_x^{(z=0)}$ = 0: the M0 in RB-I and the M1 in RB-II. Therefore, when α-MoO$_3$ is placed on gold, the fundamental mode obtains higher momentum in RB-II as it corresponds to the M1 PhP mode in an effectively twice thicker free-standing waveguide, while it becomes less confined in RB-I as it remains the first-order mode in the effectively twice thicker waveguide. This phenomenon is of key importance for the operation of PFC in RBs of different types.

The investigated one-dimensional (1D) PFC is fabricated through holographic inscription of a harmonically corrugated azopolymeric Fourier surface[28] with period $P$ = 1090 nm, followed by the deposition of a 40 nm-thick gold layer (details on sample fabrication are provided in the Supporting Information). The corrugated gold surface serves as a substrate for the pristine α-MoO$_3$ crystal, exfoliated and deposited directly onto it, as schematically shown in Fig. 2(a). As the polaritons propagate along the corrugation direction, the changing air gap size, $d$ ($d$ < 80 nm), leads to the variation of $k_P$ due to the field screening. Here, we study two PFC samples having α-MoO$_3$ thickness $t$ = 185 and 74 nm.

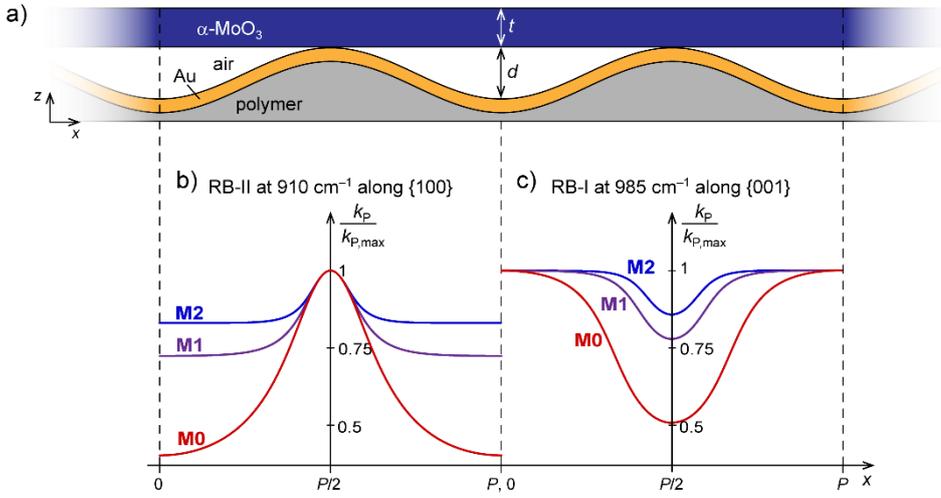

**Figure 2.** (a) Schematics of the 1D Fourier crystal based on a harmonically corrugated gold mirror and a 185 nm-thick α-MoO$_3$ slab. (b) Analytically calculated momentum of the M0, M1, and M2 PhP modes as a function of coordinate along the corrugation direction within a single unit cell, calculated at the RB-II frequency of 910 cm$^{-1}$. Momentum is normalized by its maximum at $x = P/2$ ($d = 0$) for each mode. (c) Same as in (b), calculated at the RB-I frequency of 985 cm$^{-1}$ where polaritons have anomalous dispersion and opposite field symmetry compared to RB-II. Here, momentum is normalized by its maximum at $x = 0$ ($x = P$) for each mode. Here, polariton propagation is along the {001} crystalline axis, consistently with the experiments. In both RB-I and RB-II, the momentum of M0 PhP mode is strongly modulated, while that of the higher-order modes experiences much weaker modulation depth.

Figure 2(b) shows the calculated $k_P$ variation for the fundamental, second-, and third-order modes (M0, M1, and M2, respectively) as a function of coordinate along the corrugation direction (along the lattice vector), at the frequency of 910 cm$^{-1}$ in RB-II. The higher-order modes possess higher momentum and are thus more confined, being less sensitive to the gap variation compared to the fundamental mode. Therefore, while M0 experiences ≈55% modulation of $k_P$, M1 is modulated by a mere ≈25%, and the modulation depth further decreases for higher orders. Furthermore, with a suppressed inter-mode scattering, M0 mode carries most of the energy and dominates the near-field signal, as we demonstrated in Ref. [16] for PFC with hBN. Note that $k_P$ is maximal at the contacts of α-MoO$_3$ with gold ($d = 0$). At the same time, the spatial modulation profile of $k_P$ in RB-I is inverted relative to the corrugation pattern: polaritons are least confined when $d = 0$ (Fig. 2(c); calculated at 985 cm$^{-1}$). This leads to a significantly different behavior of polaritons in PFC in RB-I which can be observed by near-field imaging.

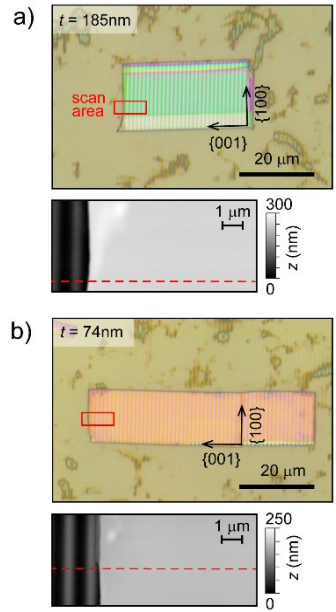

**Figure 3.** (a) Top panel: optical micrograph image of the 185 nm-thick α-MoO$_3$ flake on top of the gilded 1D Fourier surface with corrugation period of 1090 nm; real colors. Red rectangle indicates the area of the near-field imaging. Lower panel: topography of the area indicated by the red rectangle in the top panel, measured by AFM. Dashed red line indicates where near-field is measured at different frequencies. (b) Same as in (a) for the sample with 74 nm-thick α-MoO$_3$ flake. Note that even such a thin flake remains flat over the micrometer-long corrugation period.

Optical micrograph images of the samples used in this study are shown in top panels of Figs. 3(a,b), together with their topography obtained by atomic force microscopy (AFM; bottom panels). Notably, even the 74 nm-thick flake remains flat over the 1090 nm corrugation span, indicating the remarkable rigidity of α-MoO$_3$. The edges of both flakes are suspended between the surface apexes. Near-field imaging is performed over the areas indicated by the red rectangles in Fig. 3. We chose the {001} crystalline direction of α-MoO3 since phonon-polaritons in RB-I propagating along this direction are less lossy compared to {100} direction[21]. Besides, α-MoO$_3$ naturally exfoliates into crystals that are much longer in the {001} direction, which makes it much preferred for probing 1D polaritonic crystals. The near-field amplitude is measured along the red dashed lines shown in the AFM scans.

We collect the hyperspectral near-field data within the RB-I by taking individual measurements at each frequency using the tunable quantum cascade laser (MIRcat, Daylight Solutions, USA) as the excitation source for the scattering-type scanning near-field optical microscope[29] (s-SNOM; attocube systems AG, Germany). In our samples, the polaritons are primarily excited via light scattering by the material edge[30], which is suspended between the surface corrugations in both samples (AFM data in Fig. 3). The edge is always oriented orthogonal to the excitation beam in s-SNOM to maximize the scattering efficiency. The edge-launched polaritons interfere with the excitation beam and generate near-field interference fringes with the period approximately equal to the polaritonic wavelength, allowing a straightforward dispersion analysis[30]. All near-field data in this work is analyzed at the third demodulation harmonic of the near-field signal from s-SNOM, which is practically free from the background[31] (see Supporting Information for details on near-field imaging).

The hyperspectral near-field image of the thicker sample is presented in Fig. 4(a), where the AFM data is used for spatial alignment of the individual profiles so that $x = 0$ corresponds to the α-MoO$_3$ edge. The near-field amplitude from every measurement is normalized by the average value of the signal above the gold surface away from the α-MoO$_3$ flake at $x < 0$. Since the phonon-polaritons in RB-I have anomalous dispersion, the near-field fringes at lower frequencies reveal more confined polaritons with shorter $\lambda_P$. Since α-MoO$_3$ edge is suspended up to $x \approx P/2$, the signal maxima within fringes with $\lambda_P < P/2$ at lower frequencies (where PhP is more confined and thus less sensitive to screening), are expected to follow the analytically calculated $\lambda_P(\omega)$ in suspended material. Indeed, $\lambda_P(\omega)$ perfectly matches the position of the signal maximum in the first interference fringe until the frequency of 988 cm$^{-1}$ when $\lambda_P \approx P/2$ (yellow dotted line in Fig. 4(a)). The second fringe maximum is naturally fitted by the $\lambda_P$ in α-MoO$_3$ on gold which is twice larger (red dotted line in Fig. 4(a)).

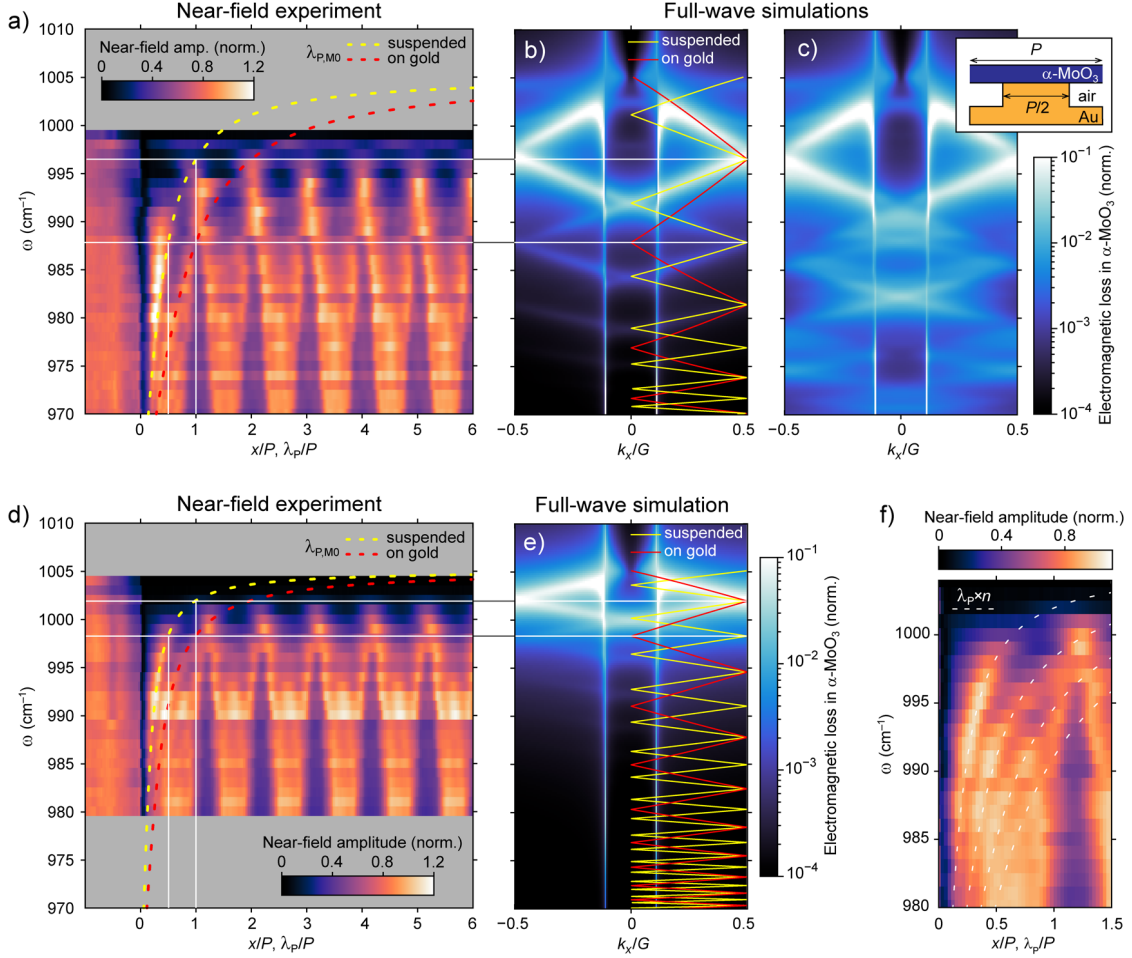

**Figure 4.** (a) Hyperspectral near-field image of the Fourier crystal with 185 nm-thick α-MoO$_3$ flake, measured at RB-I frequencies. Near-field profiles at individual frequencies are spatially aligned based on the AFM data so that $x = 0$ corresponds to the α-MoO$_3$ edge. Dotted curves show the analytically calculated wavelength of the phonon-polaritons $\lambda_P(\omega)$ in the suspended ($d = 80$ nm; yellow) and gold-supported ($d = 0$; red) flakes in a planar structure analyzed in Figs. 1(b,c). White guidelines mark the spectral boundaries of the dispersionless near-field signal associated with the Bloch mode when $P/2 < \lambda_P < P$. Near-field is normalized by its value above gold far from α-MoO$_3$. (b) Numerically calculated band diagram of the fundamental Bloch mode in the infinite 1D Fourier crystal that have sample's geometry. Solid curves show the folded dispersion of polaritons in suspended ($d = 80$ nm; yellow) and gold-supported ($d = 0$; blue) α-MoO$_3$ in the planar structure. White guidelines connecting (a) and (b) indicate the perfect agreement between the calculations and the near-field patterns associated with the Bloch mode. (c) Numerically calculated band diagram of the Bloch modes in a conventionally patterned periodic structure with sharp edges (inset), showing multiple high-order modes which are suppressed in the Fourier crystal. (d) Same as in (a) for the Fourier crystal with 74 nm-thick α-MoO$_3$ flake. (e) Same as (b) for 74 nm-thick α-MoO$_3$ sample. (f) Close-up of the hyperspectral near-field image in (d), showing perfect agreement between the calculated $\lambda_P(\omega)$ in the suspended α-MoO$_3$ (dashed lines) and the near-field fringes near its edge. Here, near-field is normalized by its maximum for each scan.

Notably, the near-field pattern over the PFC far from the edge is very similar at all frequencies where $P/2 < \lambda_P < P$ (988 cm$^{-1}$ < $\omega$ < 996 cm$^{-1}$; between the two white horizontal guidelines in Fig. 4(a)). This dispersionless near-field signal can be explained by the excitation of the fundamental Bloch mode in the structure within the first Brillouin zone. Since the collective Bloch mode has a delocalized nature, its spatial field profile within a unit cell corresponds the most energetically favorable one. In the PFC, this would correspond to the field maxima in the suspended material – precisely as seen in the experimental data in Fig. 4(a). The similar field distribution was observed in the PFC based on hBN and also associated with the presence of a Bloch mode (Fig. 5 in Ref.[16]). Interestingly, this indicates that in both RB-I and RB-II, Bloch modes have similar field distribution. It is corroborated by the observation of the near-field amplitude peaking over the suspended material. Otherwise, when the Bloch mode is not excited (e.g. within the bandgap[16]), the near-field is maximal where gold is closest to the sample surface. This is clearly visible in α-MoO$_3$ at lower frequencies where Bloch mode is damped for $\lambda_P \ll P/2$ (at $\omega \lesssim$ 980 cm$^{-1}$ in Fig. 4(a)). However, observed here hyperspectral near-field image is largely different from that in PFC with hBN.

In order to better understand the near-field patterns, we numerically calculate the band diagram of Bloch modes in the PFC, shown in Fig. 4(b) (see Supporting Information for details). First of all, we note that the band diagram is dominated by a single mode – in agreement with our previous observations in Ref.[16] where the dominant excitation of M0 PhP Bloch mode in the PFC was first demonstrated. When overlapped with the analytically calculated dispersion of an empty lattice (i.e. in the α-MoO$_3$ above a flat gold substrate at $d = 0$ and $d = 80$ nm), the Bloch mode's dispersion at higher frequencies ($\lambda_P > P$, i.e. at $\omega >$ 996 cm$^{-1}$) follows the dispersion of the M0 mode in α-MoO$_3$ on gold ($d = 0$; red curve). However, at lower frequencies ($\lambda_P < P$, i.e. at $\omega <$ 996 cm$^{-1}$), Bloch mode's dispersion precisely follows the dispersion of M0 mode in suspended α-MoO$_3$ at $d = 80$ nm, but with extra momentum of $G/2$ (yellow curve), where $G = 2\pi/P$ is the reciprocal lattice vector.

The mode-selective property of the PFC can be clearly illustrated by calculating the band diagram of the conventional binary structure with periodically patterned gold presented in Fig. 4(c). Sharp gold edges of such a structure provide the intermode scattering mechanism and thus, result in the excitation of multiple high-order Bloch modes which are as intense as the fundamental M0 PhP mode.

The second sample with 74 nm-thick α-MoO$_3$ is analyzed in the same way. The hyperspectral near-field image and the band diagram for this sample are shown in Fig. 4(d) and Fig. 4(e), respectively. In a notable similarity, despite much shorter polariton wavelengths, the same near-field features associated with the Bloch mode are observed when $P/2 < \lambda_P < P$ (998 cm$^{-1}$ $< \omega < 1002$ cm$^{-1}$; marked by two white horizontal guidelines in Fig. 4(d)). The band diagram with the analytical dispersion of M0 (Fig. 4(e)) indicates the presence of only the fundamental Bloch mode. Similar to the 185 nm-thick sample, Bloch mode bands align with the dispersion in the suspended α-MoO$_3$ for $\lambda_P < P$ ($\omega < 1002$ cm$^{-1}$; yellow curve) and with the dispersion on gold when $\lambda_P > P$ ($\omega > 1002$ cm$^{-1}$; red curve). Such a good agreement between the two samples indicates the consistent correspondence between the observed near-field pattern and the calculated polaritonic dispersion.

Finally, we highlight the good agreement between the analytically calculated $\lambda_P(\omega)$ and the near-field fringes near the edge of the suspended 74 nm-thick α-MoO$_3$ in Fig. 4(f). To improve visibility, the near-field amplitude is re-normalized by its maximum in each scan. White dashed lines show the calculated $\lambda_P(\omega)n$ for $n = 1, 2, 3$, and 4. Notably, $\lambda_P(\omega)$ agrees with the fringes up to $x \approx 0.7P$ where the gold surface gets into contact with α-MoO$_3$.

In conclusion, we have introduced PFC operating at mid-infrared frequencies for hyperbolic PhPs with anomalous dispersion in α-MoO$_3$ (in RB-I). The harmonic and mode-selective modulation of the polariton momentum in the PFC favors an excitation of the fundamental Bloch mode, while the higher-order modes are strongly suppressed. In this study, Bloch waves in α-MoO$_3$ are excited by the material edge and form a distinct dispersionless near-field distribution when $P/2 < \lambda_P < P$, consistently observed in the two samples with largely different $\lambda_P$. Particularly, the near-field maxima associated with the Bloch mode are observed over the suspended sections of the waveguide, in agreement with the calculated band diagram and the M0 dispersion in suspended α-MoO$_3$. The earlier near-field investigation of the PFC based on hBN within the RB-II revealed very different near-field response and the band diagram[16]. We attribute this to the opposite field symmetry of the M0 PhP mode in RB-I, which leads to the different momentum modulation regime in the PFC. Our results highlight the intriguing properties of the polaritons with anomalous dispersion propagating in the Fourier crystal and the polaritonic crystals in general, and demonstrate the advantages of the Fourier crystal for nanolight manipulation even in relatively lossy polaritonic crystal.

**Author Contributions**

S.G.M. conceived the research idea, fabricated the samples, performed s-SNOM measurements, performed calculations, analyzed the data with valuable input from A.Y.N. and P.A.G, and wrote the manuscript. Y.L. and H.N. fabricated the azopolymeric Fourier surfaces under the supervision of S.L. J.T.H. assisted in samples fabrication and measurements. M.S.J. supervised the project. All authors contributed to writing the manuscript vie discussions and comments.

**Notes**

The authors declare no competing financial interest.

**Acknowledgements**

This research was supported by the National Research Foundation of Korea (NRF) grants funded by the Ministry of Science, ICT and Future Planning (NRF-RS-2022-NR070476 and NRF-RS-2024-00340639). S.L. acknowledges support from NRF of Korea grants (NRF-RS-2022-NR068141 and NRF-RS-2023-00272363), the Korea-US collaborative research fund (KUCRF, RS-2024-00468463), Samsung Research Funding & Incubation Center for Future Technology (grant SRFC-MA2301-02), KU-KIST research program (2V09840-23-P023), and Korea University grant. A.Y.N. acknowledges support from the Spanish Ministry of Science and Innovation (grant PID2023-147676NB-I00) and the Basque Department of Education (grant PIBA-2023-1-0007). This work was also supported by the BK21 FOUR Program through the NRF funded by Ministry of Education.

REFERENCES

(1) Huang, K. C.; Lidorikis, E.; Jiang, X. Y.; Joannopoulos, J. D.; Nelson, K. A.; Bienstman, P.; Fan, S. H. Nature of lossy Bloch states in polaritonic photonic crystals. *Phys. Rev. B* **2004**, *69* (19), 195111.

(2) Novoselov, K. S.; Mishchenko, A.; Carvalho, A.; Neto, A. H. C. 2D materials and van der Waals heterostructures. *Science* **2016**, *353* (6298), aac9439.

(3) Basov, D. N.; Fogler, M. M.; de Abajo, F. J. G. Polaritons in van der Waals materials. *Science* **2016**, *354* (6309), aag1992.

(4) Basov, D. N.; Asenjo-Garcia, A.; Schuck, P. J.; Zhu, X. Y.; Rubio, A. Polariton panorama. *Nanophoton.* **2021**, *10* (1), 549-577.

(5) Galiffi, E.; Carini, G.; Ni, X.; Alvarez-Pérez, G.; Yves, S.; Renzi, E. M.; Nolen, R.;

Wasserroth, S.; Wolf, M.; Alonso-Gonzalez, P.; et al. Extreme light confinement and control in low-symmetry phonon-polaritonic crystals. *Nat. Rev. Mater.* **2023**, *9*, 9-28.

(6) Alonso-Gonzalez, P.; Nikitin, A. Y.; Gao, Y.; Woessner, A.; Lundeberg, M. B.; Principi, A.; Forcellini, N.; Yan, W. J.; Velez, S.; Huber, A. J.; et al. Acoustic terahertz graphene plasmons revealed by photocurrent nanoscopy. *Nat. Nanotechnol.* **2017**, *12* (1), 31-35.

(7) Xiong, L.; Forsythe, C.; Jung, M.; McLeod, A. S.; Sunku, S. S.; Shao, Y. M.; Ni, G. X.; Sternbach, A. J.; Liu, S.; Edgar, J. H.; et al. Photonic crystal for graphene plasmons. *Nat. Commun.* **2019**, *10*, 4780.

(8) Xiong, L.; Li, Y. T.; Jung, M.; Forsythe, C.; Zhang, S.; McLeod, A. S.; Dong, Y. A.; Liu, S.; Ruta, F. L.; Li, C.; et al. Programmable Bloch polaritons in graphene. *Sci. Adv.* **2021**, *7* (19), eabe8087.

(9) Yang, J.; Krix, Z. E.; Kim, S.; Tang, J. B.; Mayyas, M.; Wang, Y. F.; Watanabe, K.; Taniguchi, T.; Li, L. H.; Hamilton, A. R.; et al. Near-Field Excited Archimedean-like Tiling Patterns in Phonon-Polaritonic Crystals. *ACS Nano* **2021**, *15* (5), 9134-9142.

(10) Sheinfux, H. H.; Jung, M. W.; Orsini, L.; Ceccanti, M.; Mahalanabish, A.; Martinez-Cercós, D.; Torre, I.; Ruiz, D. B.; Janzen, E.; Edgar, J. H.; et al. Transverse Hypercrystals Formed by Periodically Modulated Phonon Polaritons. *ACS Nano* **2023**, *17* (8), 7377-7383.

(11) Alfaro-Mozaz, F. J.; Rodrigo, S. G.; Alonso-González, P.; Vélez, S.; Dolado, I.; Casanova, F.; Hueso, L. E.; Martín-Moreno, L.; Hillenbrand, R.; Nikitin, A. Y. Deeply subwavelength phonon-polaritonic crystal made of a van der Waals material. *Nat. Commun.* **2019**, *10*, 42.

(12) Alfaro-Mozaz, F. J.; Rodrigo, S. G.; Vélez, S.; Dolado, I.; Govyadinov, A.; Alonso-González, P.; Casanova, F.; Hueso, L. E.; Martín-Moreno, L.; Hillenbrand, R.; et al. Hyperspectral Nanoimaging of van der Waals Polaritonic Crystals. *Nano Lett.* **2021**, *21* (17), 7109-7115.

(13) Capote-Robayna, N.; Matveeva, O. G.; Volkov, V. S.; Alonso-González, P.; Nikitin, A. Y. Twisted Polaritonic Crystals in Thin van der Waals Slabs. *Laser Photonics Rev.* **2022**, *16* (9), 2200428.

(14) Lv, J. T.; Wu, Y. J.; Liu, J. Y.; Gong, Y. N.; Si, G. Y.; Hu, G. W.; Zhang, Q.; Zhang, Y. P.; Tang, J. X.; Fuhrer, M. S.; et al. Hyperbolic polaritonic crystals with configurable low-symmetry Bloch modes. *Nat. Commun.* **2023**, *14* (1), 3894.

(15) Sahoo, N. R.; Kumar, B.; Prasath, S. S. J.; Dixit, S.; Kumar, R.; Bapat, A.; Sharma, P.; Caldwell, J. D.; Kumar, A. Polaritons in Photonic Hypercrystals of van der Waals Materials. *Adv. Funct. Mater.* **2024**, *34* (41), 2316863.

(16) Menabde, S. G.; Lim, Y.; Voronin, K. V.; Heiden, J. T.; Nikitin, A. Y.; Lee, S.; Jang, M. S. Polaritonic Fourier crystal. *Nat. Commun.* **2025**, *16*, 2530.

(17) Dai, S.; Fei, Z.; Ma, Q.; Rodin, A. S.; Wagner, M.; McLeod, A. S.; Liu, M. K.; Gannett, W.; Regan, W.; Watanabe, K.; et al. Tunable Phonon Polaritons in Atomically Thin van der Waals Crystals of Boron Nitride. *Science* **2014**, *343* (6175), 1125-1129.

(18) Caldwell, J. D.; Aharonovich, I.; Cassabois, G.; Edgar, J. H.; Gil, B.; Basov, D. N. Photonics with hexagonal boron nitride. *Nat. Rev. Mater.* **2019**, *4* (8), 552-567.

(19) Ma, W. L.; Alonso-Gonzalez, P.; Li, S. J.; Nikitin, A. Y.; Yuan, J.; Martin-Sanchez, J.; Taboada-Gutierrez, J.; Amenabar, I.; Li, P. N.; Velez, S.; et al. In-plane anisotropic and ultra-low-loss polaritons in a natural van der Waals crystal. *Nature* **2018**, *562* (7728), 557-562.

(20) Zheng, Z. B.; Xu, N. S.; Oscurato, S. L.; Tamagnone, M.; Sun, F. S.; Jiang, Y. Z.; Ke, Y. L.; Chen, J. N.; Huang, W. C.; Wilson, W. L.; et al. A mid-infrared biaxial hyperbolic van der Waals crystal. *Sci. Adv.* **2019**, *5* (5), eaav8690.

(21) Menabde, S. G.; Jahng, J.; Boroviks, S.; Ahn, J.; Heiden, J. T.; Hwang, D.; Lee, E. S.; Mortensen, N. A.; Jang, M. S. Low-Loss Anisotropic Image Polaritons in van der Waals Crystal α-$MoO_3$. *Adv. Opt. Mater.* **2022**, *10* (21), 2201492.

(22) Alvarez-Perez, G.; Foland, T. G.; Errea, I.; Taboada-Gutierrez, J.; Duan, J. H.; Martin-Sanchez, J.; Tresguerres-Mata, A. I. F.; Matson, J. R.; Bylinkin, A.; He, M. Z.; et al. Infrared Permittivity of the Biaxial van der Waals Semiconductor α-$MoO_3$ from Near- and Far-Field Correlative Studies. *Adv. Mater.* **2020**, *32* (29), 1908176.

(23) Dong, W. K.; Qi, R. S.; Liu, T. S.; Li, Y.; Li, N.; Hua, Z.; Gao, Z. R.; Zhang, S. Y.; Liu, K. H.; Guo, J. D.; et al. Broad-Spectral-Range Sustainability and Controllable Excitation of Hyperbolic Phonon Polaritons in α-$MoO_3$. *Adv. Mater.* **2020**, *32* (46), 2002014.

(24) Passler, N. C.; Paarmann, A. Generalized 4 x 4 matrix formalism for light propagation in anisotropic stratified media: study of surface phonon polaritons in polar dielectric heterostructures. *J. Opt. Soc. Am. B* **2017**, *34* (10), 2128-2139.

(25) Alvarez-Perez, G.; Voronin, K. V.; Volkov, V. S.; Alonso-Gonzalez, P.; Nikitin, A. Y. Analytical approximations for the dispersion of electromagnetic modes in slabs of biaxial crystals. *Phys. Rev. B* **2019**, *100* (23), 235408.

(26) Lee, I. H.; He, M. Z.; Zhang, X.; Luo, Y. J.; Liu, S.; Edgar, J. H.; Wang, K.; Avouris, P.; Low, T.; Caldwell, J. D.; et al. Image polaritons in boron nitride for extreme polariton confinement with low losses. *Nat. Commun.* **2020**, *11* (1), 3649.

(27) Menabde, S. G.; Heiden, J. T.; Cox, J. D.; Mortensen, N. A.; Jang, M. S. Image polaritons in van der Waals crystals. *Nanophoton.* **2022**, *11* (11), 2433-2452.

(28) Lim, Y.; Kang, B.; Hong, S. J.; Son, H.; Im, E.; Bang, J.; Lee, S. A Field Guide to Azopolymeric Optical Fourier Surfaces and Augmented Reality. *Adv. Funct. Mater.* **2021**, *31* (39), 2104105.

(29) Hillenbrand, R.; Abate, Y.; Liu, M. K.; Chen, X. Z.; Basov, D. N. Visible-to-THz near-field nanoscopy. *Nat. Rev. Mater.* **2025**, *10*, 285-310.

(30) Jang, M.; Menabde, S. G.; Kiani, F.; Heiden, J. T.; Zenin, V. A.; Mortensen, N. A.; Tagliabue, G.; Jang, M. S. Fourier analysis of near-field patterns generated by propagating polaritons. *Phys. Rev. Appl.* **2024**, *22* (1), 014076.

(31) Ocelic, N.; Huber, A.; Hillenbrand, R. Pseudoheterodyne detection for background-free near-field spectroscopy. *Appl. Phys. Lett.* **2006**, *89* (10), 101124.

# Supporting Information

## Bloch phonon-polaritons with anomalous dispersion in polaritonic Fourier crystals

Sergey G. Menabde,[1] Yongjun Lim,[2] Alexey Y. Nikitin,[3] Pablo Alonso González,[4] Jacob T. Heiden,[1] Heerin Noh,[5] Seungwoo Lee,[2,5,6] and Min Seok Jang[1,*]

[1]School of Electrical Engineering, Korea Advanced Institute of Science and Technology, Daejeon 34141, Korea

[2]Department of Biomicrosystem Technology, Korea University, Seoul 02841, Korea

[3]Donostia International Physics Center (DIPC), Donostia-San Sebastián 20018, Spain; IKERBASQUE, Basque Foundation for Science, Bilbao 48013, Spain

[4]Department of Physics, University of Oviedo, Oviedo 33003, Spain; Center of Research on Nanomaterials and Nanotechnology, CINN (CSIC-Universidad de Oviedo), El Entrego, Spain

[5]KU-KIST Graduate School of Converging Science and Technology, Korea University, Seoul 02841, Korea

[6]Department of Integrative Energy Engineering, Korea University, Seoul 02841, Korea

*jang.minseok@kaist.ac.kr

## S1. Sample preparation

A solution of poly-dispersed red 1 methacrylate (pDR1m) with a molecular weight of 3kDa and a polydispersity index of 1.1-1.2 was prepared by dissolving it into 1,1,2-trichloroehtane solvent (Sigma-Aldrich) with 3 wt%. Solution parameters have been optimized for the most efficient development of Fourier surfaces [S1]. Then, we spin-coated the pDR1m solution onto a 2×2 cm$^2$ silicon wafer, which was cleaned through sequential sonication in acetone, isopropyl alcohol, and deionized water. The spin rate was set to 1000 rpm for 40 seconds, resulting in the azopolymeric thin film of thickness of 150 nm.

To fabricate a single-harmonic 1D Fourier surface on the azopolymeric thin film, we used an interference pattern generated by the two beams with the left- and right-handed circular polarization. In this process, we used a continuous wave diode laser (Light House Sprout) with a wavelength of 532 nm. The intensity of each beam was set to 150 mW/cm$^2$. Fabrication of the Fourier surface with a period of 485 nm and a modulation height of 70 nm required the inscription time and the beams incident angles of 10 minutes and 33 degrees, respectively.

Finally, an approximately 30 nm-thick gold layer was directly deposited onto the Fourier surface by thermal deposition in vacuum (air pressure < 5×10$^{-7}$ Torr) with deposition rate of 0.2 Å/s.

## S2. Near-field experiments

Near-field measurements were performed using the neaSNOM from attocube systems AG (formerly Neaspec) coupled with the tunable quantum cascade laser (MIRcat, Daylight Solutions). The Pt-coated AFM nanotips used in s-SNOM (ARROWNCPt, Nano World) had a typical tapping frequency $\Omega$ around 270 kHz, and the tapping amplitude was ≈70 nm in a non-contact mode. The background-free interferometric signal [S2] demodulated at the third harmonic (3$\Omega$) was used to collect near-field data.

## S3. Full-wave numerical simulations

The band structure of the PFC is visualized using the full-wave numerical simulations in the 2D domain in periodic configuration by applying the unit cell boundary conditions with Floquet periodicity. Since the gold surface is continuous, it is possible to assume the Otto excitation scheme with a fictitious high-index (*n* > 20) prism underneath the structure which provides the necessary in-plane momentum range (illustrated in Supporting Fig. S1). Then, the dispersion of Bloch phonon-polaritons in α-MoO3 and their relative intensity is revealed by the electromagnetic power loss in α-MoO3 as a function of frequency and in-plane momentum of the impinging TM-polarized plane wave in the Otto configuration.

## S3. Dielectric function of α-MoO3

Dielectric function of α-MoO3 can be modelled by a system of three Lorentzian oscillators in a general TO-LO form, providing the permittivity for each crystallographic direction:

$$\varepsilon_m(\omega) = \varepsilon_{\infty,m} \left( \frac{\omega_{LO,m}^2 - \omega^2 - i\Gamma_m \omega}{\omega_{TO,m}^2 - \omega^2 - i\Gamma_m \omega} \right), m = x, y, z$$

where $\omega_{LO}$, $\omega_{TO}$, $\Gamma$ and $\varepsilon_\infty$ are the LO and TO phonon frequencies, phonon damping, and the high-frequency permittivity, respectively. The anisotropic permittivity tensor is given by:

$$\bar{\bar{\varepsilon}} = \begin{bmatrix} \varepsilon_x & 0 & 0 \\ 0 & \varepsilon_y & 0 \\ 0 & 0 & \varepsilon_z \end{bmatrix}$$

where *x*, *y*, and *z* correspond to {001}, {100}, and {010} crystallographic axes, respectively.

In this work, we used the same α-MoO3 crystal as in our previous work [S3], which was purchased from 2D semiconductors USA. Experimentally determined parameters of the dielectric function are shown below in the Table S1, fitted from the near-field fringes of phonon-polaritons measured at different propagation directions on a low-loss monocrystalline gold substrate [S3].

|   | $\omega_{TO}$ | $\omega_{LO}$ | $\Gamma$ | $\varepsilon_\infty$ |
|---|---|---|---|---|
| *x* | 820 | 965.6 | 4 | 7.562 |
| *y* | 545 | 851 | 4 | 7.854 |
| *z* | 964 | 1005.1 | 1.3 | 3.1 |

**Table S1.** Recovered parameters of the α-MoO3 dielectric function.

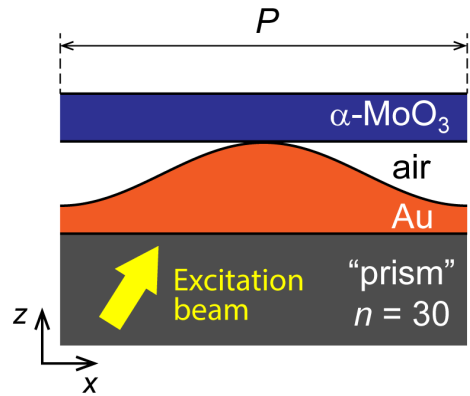

**Figure S1.** Full-wave simulation model to calculate the band structure in the polaritonic Fourier crystal. The band structure is calculated for unit cell of period $P$ by full-wave numerical simulations in 2D domain with periodic boundary conditions on both sides of the unit cell. Mapping of the electromagnetic loss density in the α-MoO$_3$ slab as a function of the excitation frequency and $x$-component of the wavevector reveals the presence of Bloch modes. The simulated excitation scheme mimics the Otto configuration where the TM-polarized plane wave illuminates the structure from the side of the substrate, which is typically a coupling prism made of a high-index material. In the simulations, the excitation wave is launched by a periodic port immersed in a fictitious high-index "prism" with $n = 30$. Such a high index is required to sweep through the necessary $k_x$ values with relatively small incidence angles < 20°. This provides a relatively uniform excitation efficiency due to the small variation of the reflection coefficient.

**References**

[S1] Lim, Y.; Kang, B.; Hong, S. J.; Son, H.; Im, E.; Bang, J.; Lee, S. A Field Guide to Azopolymeric Optical Fourier Surfaces and Augmented Reality. *Adv. Funct. Mater.* **2021**, *31*, 2104105.

[S2] Ocelic, N.; Huber, A.; Hillenbrand, R. Pseudoheterodyne detection for background-free near-field spectroscopy. *Appl. Phys. Lett.* **2006**, *89*, 101124.

[S3] Menabde, S. G.; Jahng, J.; Boroviks, S.; Ahn, J.; Heiden, J. T.; Hwang, D. K.; Lee, E. S.; Mortensen, N. A.; Jang, M. S. Low-Loss Anisotropic Image Polaritons in van der Waals Crystal α-MoO$_3$. *Adv. Opt. Mater.* **2022**, *10*, 2201492.